# Multi-step stochastic mechanism of polarization reversal in orthorhombic ferroelectrics

Y. A. Genenko[1*], M.-H. Zhang[2], I. S. Vorotiahin[1], R. Khachaturyan[3], Y.-X. Liu[4], J.-W. Li[4], K. Wang[4], and J. Koruza[2]

[1]*Materials Modelling, Department of Materials and Earth Sciences, Technical University of Darmstadt, Otto-Berndt- Str. 3, 64287 Darmstadt, Germany*

[2] *Nonmetallic Inorganic Materials, Department of Materials and Earth Sciences, Technical University of Darmstadt, Alarich-Weiss-Straße 2, 64287 Darmstadt, Germany*

[3]*Interdisciplinary Center for Advanced Materials Simulation, Ruhr-Universität Bochum, Universitätsstr. 150, 44801 Bochum, Germany*

[4]*State Key Laboratory of New Ceramics and Fine Processing, School of Materials Science and Engineering, Tsinghua University, Beijing, China*

**Abstract**

A stochastic model of electric field-driven polarization reversal in orthorhombic ferroelectrics is advanced, providing a description of their temporal electromechanical response. The theory accounts for all possible parallel and sequential switching events. Application of the model to the simultaneous measurements of polarization and strain kinetics in a lead-free orthorhombic $(K,Na)NbO_3$-based ferroelectric ceramic over a wide timescale of 7 orders of magnitude allowed identification of preferable polarization switching paths, fractions of individual switching processes, and their activation fields. Particularly, the analysis revealed substantial contributions of coherent non-180° switching events, which do not cause macroscopic strain and thus mimic 180° switching processes.

[*] Corresponding author: genenko@mm.tu-darmstadt.de

1. Introduction

Ferroelectric materials find numerous applications as transducers and actuators, electromechanical and infrared image sensors, ferroelectric nonvolatile memories (FeRAM), etc. [1]. Their constitutive characteristic is the appearance, below a transition temperature, of a spontaneous polarization, which can be redirected (switched) by application of electric field. Fundamental physical properties of ferroelectrics, such as the remanent polarization [2], permittivity [3] and piezoelectric coefficient [4] are strongly dependent on the crystallographic structure and can be modified by composition [2,5] and application of mechanical stresses [6,7]. For such applications as FeRAM, fast polarization switching is crucial since it allows high operating frequencies of ferroelectric devices. The field-dependent switching times are also strongly dependent on the crystallographic structure and may be controlled by composition [8-10]. Particularly, optimization of the switching kinetics is significant for the prospective multi-level data storage [11-14]. However, the understanding of the underlying mechanisms is still incomplete [15-17], especially because, for most ferroelectrics, polarization switching involves complicated ferroelastic switching processes [18].

The stochastic Kolmogorov-Avrami-Ishibashi (KAI) model roughly captures the main features of the polarization switching kinetics in ferroelectrics. This commonly used approach, originally advanced to describe melt solidification [19], assumes random and statistically-independent nucleation and growth of reversed polarized domains in a uniformly polarized medium when an opposite electric field is applied [20,21]. Being initially developed for a scalar order parameter, the KAI model considers variation of only one polarization component. Thus, ferroelastic switching processes in multiaxial ferroelectrics, which involve a few polarization components, cannot be properly accounted for. Another related feature that is missing in this approach is that the polarization reversal in reality may proceed by sequential switching events in addition to parallel events which are assumed the KAI model. Experimentally, consecutive

non-180°-switching processes were observed earlier by X-ray diffraction [22] and ultrasonic measurements [23]. Recently, the characteristic times of two consecutive non-180° switching steps were resolved by *in situ* X-ray diffraction measurements [24]. However, the identified fractions of 180° and non-180° processes remain controversial in literature ranging from scenarios with exclusively non-180° switching events [14,25] to those with partial non-180° processes [22-24,26,27], predominantly 180° processes [28,29] or exclusively 180° switching events [30]. Very recently, it was shown that these fractions could be substantially affected by the composition [16].

A deeper insight into the complicated switching mechanisms in ferroelectrics/ferroelastics was gained by simultaneous measurements of the switched polarization and the macroscopic strain supported by *in situ* X-ray diffraction experiments performed on a polycrystalline ferroelectric lead zirconate titanate (PZT) of tetragonal symmetry [31]. Later the simultaneous polarization and the strain measurements were also performed on PZT ceramics of rhombohedral symmetry [32] and $(K,Na)NbO_3$ (KNN)-based solid solutions [33]. An advanced quantitative analysis of the above-mentioned experimental data was performed by means of an original multistep stochastic mechanism (MSM) model developed first for the systems of tetragonal symmetry [34]. The MSM model allowed identification of the fraction of ferroelastically-active 90°-switching events. A combination of the MSM model with the inhomogeneous field mechanism (IFM) model [35,36], accounting for the distribution of the switching times due to the spatial distribution of the electric field, allowed the description of the simultaneous polarization and strain responses of tetragonal ferroelectric ceramics over a broad time range with high accuracy [37]. Further extension of the MSM model to ferroelectrics of rhombohedral crystallographic symmetry applied to the available experimental data [32] allowed extraction of fractions of 109°-, 71°- and 180°-switching events as well as their activation fields [38].

No stochastic model is available yet for description of the time-dependent electromechanical response of orthorhombic ferroelectric/ferroelastic materials where possible 60°-, 90°-, 120°- and 180°-switching events must be accounted for. At the same time, kinetics of the polarization switching in orthorhombic materials is of great interest, in particular, because of the highest polarizations and switching rates observed on $Ba(Zr_{0.2}Ti_{0.8})O_{3-x}$-$(Ba_{0.7}Ca_{0.3})TiO_3$ solid solutions of orthorhombic symmetry [9,10]. In addition, prospective electronic materials based on $HfO_2$ thin films [39] and promising piezoelectric KNN-based solid solutions [33] demonstrate ferroelectric properties in orthorhombic phases.

The current work advances a stochastic model for the kinetics of multistep switching processes during the polarization reversal in orthorhombic ferroelectrics/ferroelastics. The paper is organized as follows: In Section 2, the MSM model for single crystalline and the hybrid MSM-IFM model for polycrystalline orthorhombic systems are developed, including all possible sequential non-180° polarization reorientation steps and parallel 180° switching events. To describe the simultaneous strain kinetics, a relation between the time-dependent strain and polarization is derived. In Section 3 materials preparation and characterization are described. In Section 4, the MSM-IFM model is applied to the original polarization and strain measurements on an orthorhombic KNN-based solid solution over a time domain from $10^{-6}$ s to $10^1$ s for a range of applied electric field values ($0.3E_C < E_a < 2.6\ E_C$) around the coercive field $E_C$, and the physical meaning of parameters resulting from fitting is discussed. Finally, the results are concluded in Section 5.

## 2. Theory of consecutive stochastic polarization switching processes

In an orthorhombic ferroelectric, local polarization may adopt one of the twelve possible directions along the medial plane diagonals of the pseudo-cubic cell (<011> directions).

Application of an electric field opposite to the initial polarization direction promotes reversal of the polarization, which can proceed through different intermediate switching events. In this Section, we first display possible switching paths in a single crystalline case and then evaluate their probabilities by an extension of the KAI approach. In the following, we additionally account for the statistical distribution of the local switching times characteristic of polycrystalline materials. Hereinafter, we derive the variation of strain related to the changing polarization.

### A. Polarization rotations and switching channels

Consider a crystalline unit cell of an orthorhombic ferroelectric. Axes of a Cartesian frame are chosen to be collinear with main axes of a pseudo-cubic cell (Fig. 1(a)). The system is assumed to be initially in the polarization state $P_s(0, -1, -1)/\sqrt{2}$ (polarization pointing to A) with spontaneous polarization $P_s$ and then it is driven to the final state $P_s(0,1,1)/\sqrt{2}$ (polarization pointing to D) by application of the electric field $\boldsymbol{E}$ pointing in the $[011]$ direction. Polarization reversal may proceed along different paths exemplarily shown in Fig. 1(a), namely, by a direct 180° polarization reversal path A-D, by two consecutive 120°- and 60°- polarization rotations A-C-D, by two consecutive 60°- and 120°-polarization rotations A-B-D and by a triple consecutive 60°- polarization rotation A-B-C-D. It should be noted that here the term "rotation" is not used in the same sense as in monoclinic ferroelectrics where the polar vector can freely rotate in a given crystallographic plane, but instead refers to a rotation path through a sequence of fixed polarization directions.

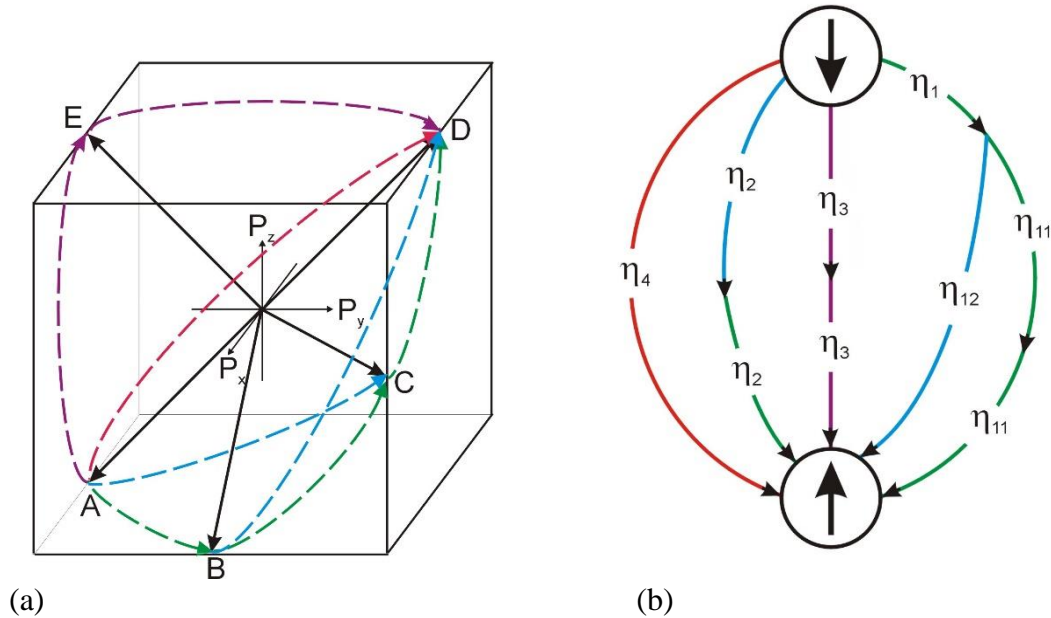

(a) (b)

Fig. 1. (a) Possible ways of the field-driven polarization reversal in orthorhombic state: exemplary 60°-60°-60° polarization rotation path A-B-C-D, exemplary 60°-120° polarization rotation path A-B-D, exemplary 120°-60° polarization rotation path A-C-D, exemplary 90°-90° polarization rotation path A-E-D, and a direct 180° polarization reversal A-D. (b) Definition of fractions of different polarization switching paths. Both in (a) and (b) 60°-switching processes are indicated with green lines, 120°-switching processes with blue lines, 90°-switching processes with purple lines and 180°-switching processes with red lines. Black arrows indicate the end of each process.

We define the probabilities for the first switching events as $\eta_1$, $\eta_2$, $\eta_3$ and $\eta_4$ for switching starting with 60°, 120°, 90° and 180° rotations, respectively, i.e. $\eta_1+\eta_2+\eta_3+\eta_4 =1$. After the first 60°-polarization rotation, the switching channel 1 splits up in two possible paths, a single 120°-polarization rotation with a weight $\eta_{12}$ and a double 60°-polarization rotation with a weight $\eta_{11}$, satisfying $\eta_{11}+\eta_{12} = \eta_1$. In a ferroelectric, switching probabilities are physically determined by the energy barriers between different polarization states, the strength and orientation of the local electric field and/or mechanical stress, as well as electric and elastic interactions between domains and boundary conditions. In a stochastic theory of the macroscopic response, the parameters $\eta_{11}$, $\eta_{12}$, $\eta_2$, $\eta_3$ and $\eta_4$ are understood as average characteristics of the whole system.

### B. Polarization variation due to multi-step switching events in uniform media (MSM model)

The classical KAI approach [19-21] assumes statistically independent switching processes at different locations and time moments. This obviously does not apply to consecutive switching processes, for which some switching events may only occur after previous ones are accomplished. The assumption of statistical independence of switching processes at different locations is also arguable. In this respect, piezoresponse force microscopy (PFM) and transmission electron microscopy (TEM) disclosed clustering during polarization reversal which ranges from a few [40] to $10^2$–$10^3$ grains [41,42] in polycrystalline thin films. In contrast to the latter results, in bulk ceramics the grain-resolved three-dimensional X-ray diffraction revealed a collective dynamics at a scale of about 10–20 grains [43-45] that roughly corresponds to the next neighbors. The study of polarization and electric field correlations by means of the self-consistent mesoscopic switching model (SMS) [46-48], revealed that the electric field-mediated correlations in bulk ceramics have a typical scale of a grain size. Physical mechanism of the short-range correlations consists in the effective screening of depolarization fields by adapting bound charges at grain boundaries. On the other hand, correlations due to elastic interactions remain an open question [31]. We keep in the following the simplified KAI assumption that switching processes at different places are statistically independent and thus neglect the electric and elastic interactions of domains during their switching to concentrate on the sequential switching events. Now we derive the probabilities of different consecutive switching steps following the MSM approach [34,38].

At the beginning, we assume a monodomain single crystal with saturation polarization along $[0\bar{1}\bar{1}]$ (see Fig. 1(a)). When an opposite electric field is applied, the local polarizations may undergo different switching processes with distinct nucleation rates per unit time and unit volume and develop growing switched domains of different geometries. In the spirit of the KAI model [19-21], we consider first the reversed domains according to the 180°-switching process A-D (Fig. 1(a)), which appear in response to the application of electric field pulse at time $t_0 = 0$. An essential element of this approach, advanced by Ishibashi and Takagi [21], is the probability $q_{180}(t,t_0)$ for an arbitrary point in space not to be comprised by a switched region of some domain,

$$q_{180}(t,t_0) = \exp\left[-\left(\frac{t-t_0}{\tau_4}\right)^{\beta_4}\right] \tag{1}$$

where $\tau_4$ is the characteristic time of 180°-switching process and $\beta_4$ is the so called Avrami index related to the dimensionality $D$ of the growing domain. Ishibashi and Takagi [21] suggested two distinct scenarios of the reversed domain nucleation: (I) the nucleation rate is constant in space and time throughout the total polarization reversal while the constant uniform electric field is held, and (II) there are only initial nuclei but no new nucleation. Assuming the velocity of domain growth to be constant (though field dependent), $\beta_4 = D+1$ applies for the regime I, and thus $\beta_4$ is larger than unity, while for the regime II, $\beta_4 = D$. This provides a contribution to the total switched polarization proportional to the fraction $(1-q_{180}(t,t_0))$ of the system volume, which reads

$$\Delta P_{180}(t) = \eta_4 2P_s L_4(t) = \eta_4 2P_s \left\{1 - \exp\left[-\left(\frac{t}{\tau_4}\right)^{\beta_4}\right]\right\} \tag{2}$$

The MSM model extends the KAI approach by applying the formula (1) to the first and the following switching events of the two-step or three-step switching processes with their respective characteristic times and Avrami indices. Similar to tetragonal ferroelectrics, in orthorhombic systems the 90°-90° polarization rotation sequence is possible, as presented exemplarily by the polarization rotation path A-E-D (Fig. 1(a)). Thus, as for the tetragonal case [34], we introduce a probability $q_{90,1}(t_1,t_0)$ not to switch according to the first 90°-switching mechanism by some intermediate time $t_1$ for the 90°-90° polarization rotation path in orthorhombic counterparts. Furthermore, we introduce a probability $q_{90,2}(t,t_1)$ not to switch according to the second 90°-switching mechanism by the time $t$ after the first switching event occurred at time $t_1$. When summarizing over all possible intermediate times $t_1$ between 0 and $t$, the total probability to switch once by the first 90°-mechanism and not to switch anymore by the second 90°-mechanism by the time $t$ is obtained as

$$L_{3\bar{3}}(t) = \frac{\beta_3}{\tau_3} \int_0^t dt_1 \left(\frac{t_1}{\tau_3}\right)^{\beta_3 - 1} \exp\left[-\left(\frac{t_1}{\tau_3}\right)^{\beta_3}\right] \exp\left[-\left(\frac{t - t_1}{\tau_{33}}\right)^{\beta_{33}}\right] \quad (3)$$

where $\tau_3$ and $\tau_{33}$ are the characteristic times of the first and the second 90°-processes, respectively, and $\beta_3$ and $\beta_{33}$ are the respective Avrami indices. By the derivation of Eq. (3), time independent nucleation rates and constant velocities of the domain growth were assumed [21]. Accordingly, the total probability to switch firstly by the first 90°-mechanism and then by the second 90°-mechanism by the time $t$ reads

$$L_{33}(t) = \frac{\beta_3}{\tau_3} \int_0^t dt_1 \left(\frac{t_1}{\tau_3}\right)^{\beta_3 - 1} \exp\left[-\left(\frac{t_1}{\tau_3}\right)^{\beta_3}\right] \left\{1 - \exp\left[-\left(\frac{t - t_1}{\tau_{33}}\right)^{\beta_{33}}\right]\right\}$$

$$= 1 - \exp\left[-\left(\frac{t}{\tau_3}\right)^{\beta_3}\right] - L_{3\bar{3}}(t). \quad (4)$$

This provides a contribution to the total switched polarization

$$\Delta P_{90-90}(t) = \eta_3 \left[ P_s L_{3\bar{3}}(t) + 2 P_s L_{33}(t) \right] \quad (5)$$

since each 90°-polarization rotation provides variation by $P_s$ along the $[011]$ direction (see Fig. 1(a)).

Considering in the same manner the 120°-60° polarization rotation path, formally similar to the 109°-71° process in rhombohedral ferroelectrics [38], we introduce a probability $q_{120}(t_1, t_0)$ not to switch according to the first 120°-switching mechanism by some intermediate time $t_1$. Furthermore, we introduce a probability $q_{60}(t, t_1)$ not to switch according to the second 60°-switching mechanism by the time $t$ after the first 120°-switching event occurred at time $t_1$.

When summarizing over all possible intermediate times $t_1$ between 0 and $t$, the total probability to switch once according to the first 120°-mechanism and not to switch anymore according to the second 60°-mechanism by the time $t$ is obtained by integration over all intermediate times $t_1$,

$$L_{2\bar{1}}(t) = \frac{\beta_2}{\tau_2} \int_0^t dt_1 \left(\frac{t_1}{\tau_2}\right)^{\beta_2 - 1} \exp\left[-\left(\frac{t_1}{\tau_2}\right)^{\beta_2}\right] \exp\left[-\left(\frac{t-t_1}{\tau_{21}}\right)^{\beta_{21}}\right] \tag{6}$$

where $\tau_2$ and $\tau_{21}$ are the characteristic times of the first 120°- and the second 60°-processes, respectively, and $\beta_2$ and $\beta_{21}$ are the respective Avrami indices. Accordingly, the total probability to switch firstly according to the first 120°-mechanism and secondly according to the second 60°-mechanism by the time $t$ reads as

$$\begin{aligned} L_{21}(t) &= \frac{\beta_2}{\tau_2} \int_0^t dt_1 \left(\frac{t_1}{\tau_2}\right)^{\beta_2 - 1} \exp\left[-\left(\frac{t_1}{\tau_2}\right)^{\beta_2}\right] \left\{1 - \exp\left[-\left(\frac{t-t_1}{\tau_{21}}\right)^{\beta_{21}}\right]\right\} \\ &= 1 - \exp\left[-\left(\frac{t}{\tau_2}\right)^{\beta_2}\right] - L_{2\bar{1}}(t). \end{aligned} \tag{7}$$

This provides a contribution to the total switched polarization

$$\Delta P_{120-60}(t) = \eta_2 \left[\frac{3}{2} P_s L_{2\bar{1}}(t) + 2 P_s L_{21}(t)\right] \tag{8}$$

since each 120°-polarization rotation provides variation by $3P_s/2$ and the following 60°-polarization rotation provides variation by $P_s/2$ along the $[011]$ direction (see Fig. 1(a)).

In the case of the first 60°-switching event, description of polarization reversal becomes more complicated because it allows splitting in two channels as depicted in Fig. 1(b). The probabilities for the two-step 60°-120° channel are essentially similar to those by the 71°-109° process in rhombohedral ferroelectrics [38] and can be described by formulas similar to Eqs.

(6,7). Namely, the total probability to switch once according to the first 60°-mechanism and not to switch anymore by the second 120°-rotation reads

$$L_{1\bar{2}}(t) = \frac{\beta_1}{\tau_1} \int_0^t dt_1 \left(\frac{t_1}{\tau_1}\right)^{\beta_1 - 1} \exp\left[-\left(\frac{t_1}{\tau_1}\right)^{\beta_1}\right] \exp\left[-\left(\frac{t - t_1}{\tau_{12}}\right)^{\beta_{12}}\right], \qquad (9)$$

where $\tau_1$ and $\tau_{12}$ are the characteristic times of the first 60°-and the second 120°- processes, respectively, and $\beta_1$ and $\beta_{12}$ are the respective Avrami indices. The total probability to switch firstly by 60° and secondly by 120° by the time $t$ reads as

$$L_{12}(t) = \frac{\beta_1}{\tau_1} \int_0^t dt_1 \left(\frac{t_1}{\tau_1}\right)^{\beta_1 - 1} \exp\left[-\left(\frac{t_1}{\tau_1}\right)^{\beta_1}\right] \left\{1 - \exp\left[-\left(\frac{t - t_1}{\tau_{12}}\right)^{\beta_{12}}\right]\right\}$$
$$= 1 - \exp\left[-\left(\frac{t_1}{\tau_1}\right)^{\beta_1}\right] - L_{1\bar{2}}(t). \qquad (10)$$

We note that the characteristic times and Avrami indices may differ between 60°-120° and 120°-60° switching since the energy barriers for both rotation angles depend on the initial polarization and strain configurations. The 60°-120° switching path provides a contribution to the total switched polarization

$$\Delta P_{60-120}(t) = \eta_{12} \left[\frac{1}{2} P_s L_{1\bar{2}}(t) + 2 P_s L_{12}(t)\right] \qquad (11)$$

since the first 60°-polarization rotation provides variation by $P_s/2$ and the following 120°-rotation provides variation by $3P_s/2$ along the $[011]$ direction (see Fig. 1(a)).

A succession of three sequential 60°- switching processes (such as the A-B-C-D path in Fig. 1(a)) is analogue to a succession of three 71°- switching events in rhombohedral ferroelectrics [38] and can be considered in a way similar to the two step processes, however, with two intermediate times $t_1$ and $t_2$ for the second and the third event, respectively. When integrating over all possible intermediate times, the probability to switch first time on the 60°-60°-60° path and not to switch anymore reads

$$L_{1\bar{1}}(t) = \frac{\beta_1}{\tau_1} \int_0^t dt_1 \left(\frac{t_1}{\tau_1}\right)^{\beta_1 - 1} \exp\left[-\left(\frac{t_1}{\tau_1}\right)^{\beta_1}\right] \exp\left[-\left(\frac{t-t_1}{\tau_{11}}\right)^{\beta_{11}}\right], \quad (12)$$

where $\tau_1$ and $\tau_{11}$ are the characteristic times of the first 60°- and the second 60°-processes, respectively, and $\beta_1$ and $\beta_{11}$ are the respective Avrami indices. Similarly, the probability to switch the first and the second time on the 60°-60°-60° path and not to switch anymore is

$$L_{11\bar{1}}(t) = \frac{\beta_1}{\tau_1} \int_0^t dt_1 \left(\frac{t_1}{\tau_1}\right)^{\beta_1 - 1} \exp\left[-\left(\frac{t_1}{\tau_1}\right)^{\beta_1}\right] \int_{t_1}^t dt_2 \frac{\beta_{11}}{\tau_{11}} \left(\frac{t_2 - t_1}{\tau_{11}}\right)^{\beta_{11} - 1} \exp\left[-\left(\frac{t_2 - t_1}{\tau_{11}}\right)^{\beta_{11}}\right]$$
$$\times \exp\left[-\left(\frac{t - t_2}{\tau_{111}}\right)^{\beta_{111}}\right] \quad (13)$$

where $\tau_{111}$ and $\beta_{111}$ are the characteristic time and the respective Avrami index of the third 60°-switching event. Finally, the probability to switch sequentially three times by 60° equals

$$L_{111}(t) = \frac{\beta_1}{\tau_1} \int_0^t dt_1 \left(\frac{t_1}{\tau_1}\right)^{\beta_1 - 1} \exp\left[-\left(\frac{t_1}{\tau_1}\right)^{\beta_1}\right] \int_{t_1}^t dt_2 \frac{\beta_{11}}{\tau_{11}} \left(\frac{t_2 - t_1}{\tau_{11}}\right)^{\beta_{11} - 1} \exp\left[-\left(\frac{t_2 - t_1}{\tau_{11}}\right)^{\beta_{11}}\right]$$
$$\times \left\{1 - \exp\left[-\left(\frac{t - t_2}{\tau_{111}}\right)^{\beta_{111}}\right]\right\} = 1 - \exp\left[-\left(\frac{t}{\tau_1}\right)^{\beta_1}\right] - L_{1\bar{1}}(t) - L_{11\bar{1}}(t) \quad (14)$$

The contribution to the total switched polarization for the path 60°-60°-60° is thus

$$\Delta P_{60-60-60}(t) = \eta_{11} \left[\frac{1}{2} P_s L_{1\bar{1}}(t) + \frac{3}{2} P_s L_{11\bar{1}}(t) + 2 P_s L_{111}(t)\right] \quad (15)$$

since the first and the third 60°-polarization rotations provide variations by $P_s/2$ and the second 60°-rotation provides variation by $P_s$ along the $[01\bar{1}]$ direction (see Fig. 1(a)).

The contributions of all switching channels sum up to a total spontaneous polarization variation of

$$\Delta P(t) = \eta_4 2P_s \left\{ 1 - \exp\left[ -\left(\frac{t}{\tau_4}\right)^{\beta_4} \right] \right\} + \eta_3 P_s \left[ L_{3\bar{3}}(t) + 2L_{33}(t) \right] + \eta_2 P_s \left[ \frac{3}{2} L_{2\bar{1}}(t) + 2L_{21}(t) \right]$$
$$+ \eta_{12} P_s \left[ \frac{1}{2} L_{1\bar{2}}(t) + 2L_{12}(t) \right] + \eta_{11} P_s \left[ \frac{1}{2} L_{1\bar{1}}(t) + \frac{3}{2} L_{11\bar{1}}(t) + 2L_{111}(t) \right]. \tag{16}$$

We note that the time dependences of probabilities $L_i(t)$ do not have a physical meaning of a gradual transformation of each particular ferroelectric cell. In line with the stochastic KAI concept, these probabilities denote the parts of the system, comprised by the respective switched domains. The local switching at a certain place is assumed to occur instantaneously as soon as this location is covered by a switched region of a growing domain.

The integrals in Eq. (16) cannot be solved in a closed form for arbitrary parameters involved. However, the functional dependences $L_i(t)$ are qualitatively similar for different values of parameters $\beta_i$. It is therefore illustrative to evaluate these dependencies exemplarily in the closed form, which is possible by setting all Avrami indices $\beta_i$ equal to unity, in order to visualize their behavior. In this case,

$$L_{3\bar{3}}(t) = \frac{\tau_{33}}{\tau_3 - \tau_{33}} \left( e^{-t/\tau_3} - e^{-t/\tau_{33}} \right) \tag{17a}$$

$$L_{33}(t) = 1 - e^{-t/\tau_3} - \frac{\tau_{33}}{\tau_3 - \tau_{33}} \left( e^{-t/\tau_3} - e^{-t/\tau_{33}} \right) \tag{17b}$$

$$L_{2\bar{1}}(t) = \frac{\tau_{21}}{\tau_2 - \tau_{21}} \left( e^{-t/\tau_2} - e^{-t/\tau_{21}} \right), \tag{17c}$$

$$L_{21}(t) = 1 - e^{-t/\tau_2} - \frac{\tau_{21}}{\tau_2 - \tau_{21}} \left( e^{-t/\tau_2} - e^{-t/\tau_{21}} \right), \tag{17d}$$

$$L_{1\bar{2}}(t) = \frac{\tau_{12}}{\tau_1 - \tau_{12}} \left( e^{-t/\tau_1} - e^{-t/\tau_{12}} \right), \tag{17e}$$

$$L_{12}(t) = 1 - e^{-t/\tau_1} - \frac{\tau_{12}}{\tau_1 - \tau_{12}} \left( e^{-t/\tau_1} - e^{-t/\tau_{12}} \right), \tag{17f}$$

$$L_{1\bar{1}}(t) = \frac{\tau_{11}}{\tau_1 - \tau_{11}} \left( e^{-t/\tau_1} - e^{-t/\tau_{11}} \right), \tag{17g}$$

$$L_{11\bar{1}}(t) = \frac{\tau_{111}}{\tau_{11} - \tau_{111}} \left[ \frac{\tau_{11}}{\tau_{11} - \tau_1} \left( e^{-t/\tau_{11}} - e^{-t/\tau_1} \right) - \frac{\tau_{111}}{\tau_{111} - \tau_1} \left( e^{-t/\tau_{111}} - e^{-t/\tau_1} \right) \right], \tag{17h}$$

$$L_{111}(t) = 1 - e^{-t/\tau_1} - \frac{\tau_{11}^2}{(\tau_{11} - \tau_{111})(\tau_{11} - \tau_1)} \left( e^{-t/\tau_{11}} - e^{-t/\tau_1} \right) + \frac{\tau_{111}^2}{(\tau_{11} - \tau_{111})(\tau_{111} - \tau_1)} \left( e^{-t/\tau_{111}} - e^{-t/\tau_1} \right). \tag{17i}$$

In the following, we do not choose *a priori* values of $\beta_i$ or other fitting parameters, but instead use them to fit the experimental data.

### C. Consecutive switching events in polycrystalline materials: A hybrid multistep-inhomogeneous field mechanism (MSM-IFM) model

Ferroelectric ceramics often exhibit a rather dispesive response to the applied electric field indicative of the statistical distribution of the local switching times, which presents a key hypothesis of the nucleation limited switching (NLS) model [49]. Therefore we advance in this section a model accounting for the distribution of the switching times in conjunction with the above presented MSM model. Since the switching time strongly depends on the electric field, the spatial distribution of local electric field values in a spatially random system like a ferroelectric ceramic may be a physical reason of the statistical distribution of the local switching times [50,51]. Assuming that the field dependence of switching times of the above introduced switching processes follows the Merz law [52], $\tau_n(E) = \tau_0 \exp\left[ (E_A^{(n)} / E)^\alpha \right]$, where $E_A^{(n)}$ is an activation field of the respective switching process and $E$ the local field value, the statistical distribution of switching times can be derived from the statistical distribution of local field values $Z(E)$ around the applied field value $E_a$. In the following this distribution will be assumed in the Lorentzian form

$$Z(E) = \frac{A}{\pi E_a} \frac{\kappa}{\left(E/E_a - 1\right)^2 + \kappa^2} \tag{18}$$

with a dimensionless width $\kappa$ which proved to work well in conjunction with the NLS model in thin polycrystalline PZT ferroelectric films [50,51]. The $\kappa$-dependent normalization constant $A = \left[1/2 + (1/\pi)\arctan(1/\kappa)\right]^{-1}$ apporaches unity when $\kappa \to 0$. A scaling form of Eq. (18) reflects the scaling properties of the polarization response observed on various bulk polycrystalline ferroelectrics [8,9,10,34-37].

Averaging over the statistical distribution of the local switching times $\tau_n$ is associated with the averaing over the statistical distribution of the local field values $E$ by a relation

$$\int_0^\infty dE\, Z(E)... = \int_{\tau_0}^\infty d\tau_n \frac{A}{\alpha\pi \cdot \tau_n} \frac{W^{(n)} \cdot \left(\ln(\tau_n/\tau_0)\right)^{1/\alpha - 1}}{\left[\left(\ln(\tau_n/\tau_0)\right)^{1/\alpha} - \left(\ln(t^{(n)}/\tau_0)\right)^{1/\alpha}\right]^2 + \left(W^{(n)}\right)^2}... \tag{19}$$

with the characteristic times $t^{(n)} = \tau_0 \exp\left[\left(\frac{E_A^{(n)}}{E_a(1+\kappa^2)}\right)^\alpha\right]$ and the distribution widths $W^{(n)} = \frac{E_A^{(n)}}{E_a}\frac{\kappa}{1+\kappa^2}$. When averaging the expressions (2-4,6,7,9,10,12-14) over the switching time distribution (19) some simplifications will be undertaken as follows. The modelling of multi-step polarization reversal in tetragonal [34,37] and rhombohedral [38] ferroelectric ceramics revealed that the first switching steps, supported by residual strains in a highly poled initial state, are typically characterized by Avrami indices $\beta$ in the range of 0.1-0.3. As was shown in Ref. [36], when an Avrami index is remakably smaller than unity and the statistical field distribution is narrow enough (that means $\kappa \ll 1$ in the context of Eq. (18)), the averaging over the switching times (19) hardly changes the Debye-like functions of the type of Eq. (2). That is why in the following the statistical averaging will not be applied to the switching times $\tau_1, \tau_2$ and $\tau_3$, related to the first 60°, 120° and 90° processes, respectively, which will be taken at the value of the applied electric field as $\tau_n = \tau_0 \exp\left[(E_A^{(n)}/E_a)^\alpha\right]$. In contrast, for the Avrami indices $\beta$ larger than unity, a Debye-like function on the logarithmic time scale can be approximated by a theta function [36], $1 - \exp\left[-(t/\tau)^\beta\right] \cong \theta(t - \tau)$, and further averaged using the distribution (19); the approximation applied for the other swithing events. Furthermore, we note that the third switching event in the 60°-60°-60° swithing path and the second switching event in the 120°-60° switching path are identical, and thus their activation energies $E_A^{(111)} = E_A^{(21)}$ and the respective switching times $\tau_{111} = \tau_{21}$. Finally, beyond the switching times $\tau_1, \tau_2$ and $\tau_3$ and the distribution width $\kappa$, the other independent parameters used for fitting the experiment in the following are the distribution widths $W^{(11)}, W^{(111)}, W^{(12)}, W^{(33)}$ and $W^{(4)}$ and the characteristic times $t^{(11)}, t^{(111)}, t^{(12)}, t^{(33)}$ and $t^{(4)}$, related to the activation fields $E_A^{(11)}, E_A^{(111)}, E_A^{(12)}, E_A^{(33)}$ and $E_A^{(4)}$, respectively, as well as the process fractions $\eta_i$. Using the introduced simplifications and performing the integration over switching times, the probability expressions (2-4,6,7,9,10,12-14) can now be evaluated for polycrystalline media.

For the 180°-switching process, the switching probability after averaging reads

$$L_4(t) \Rightarrow \frac{A}{\pi}\left\{\arctan\left[\frac{\left(\ln(t/\tau_0)\right)^{1/\alpha} - \left(\ln(t^{(4)}/\tau_0)\right)^{1/\alpha}}{W^{(4)}}\right] + \arctan\left[\frac{\left(\ln(t^{(4)}/\tau_0)\right)^{1/\alpha}}{W^{(4)}}\right]\right\}. \quad (20)$$

For the 90°-90°-switching process, the above-defined switching probabilities after averaging read

$$L_{3\bar{3}}(t) \Rightarrow A\frac{\beta_3}{\tau_3}\int_0^t dt_1 \left(\frac{t_1}{\tau_3}\right)^{\beta_3-1} \exp\left[-\left(\frac{t_1}{\tau_3}\right)^{\beta_3}\right]\left\{\frac{1}{2}-\frac{1}{\pi}\arctan\left[\frac{\left(\ln((t-t_1)/\tau_0)\right)^{1/\alpha}-\left(\ln(t^{(33)}/\tau_0)\right)^{1/\alpha}}{W^{(33)}}\right]\right\},$$

(21)

and $L_{33}(t) \Rightarrow 1-\exp\left[-\left(\frac{t}{\tau_3}\right)^{\beta_3}\right]-L_{3\bar{3}}(t).$  (22)

For the 120°-60°-switching process the above-defined switching probabilities after averaging read (note that $\tau_{111}=\tau_{21}$ and $W_{111}=W_{21}$)

$$L_{2\bar{1}}(t) \Rightarrow A\frac{\beta_2}{\tau_2}\int_0^t dt_1 \left(\frac{t_1}{\tau_2}\right)^{\beta_2-1} \exp\left[-\left(\frac{t_1}{\tau_2}\right)^{\beta_2}\right]\left\{\frac{1}{2}-\frac{1}{\pi}\arctan\left[\frac{\left(\ln((t-t_1)/\tau_0)\right)^{1/\alpha}-\left(\ln(t^{(21)}/\tau_0)\right)^{1/\alpha}}{W^{(21)}}\right]\right\}$$

(23)

and $L_{21}(t) \Rightarrow 1-\exp\left[-\left(\frac{t}{\tau_2}\right)^{\beta_2}\right]-L_{2\bar{1}}(t).$  (24)

For the 60°-120°-switching process the above-defined switching probabilities after averaging read

$$L_{1\bar{2}}(t) \Rightarrow A\frac{\beta_1}{\tau_1}\int_0^t dt_1 \left(\frac{t_1}{\tau_1}\right)^{\beta_1-1} \exp\left[-\left(\frac{t_1}{\tau_1}\right)^{\beta_1}\right]\left\{\frac{1}{2}-\frac{1}{\pi}\arctan\left[\frac{\left(\ln((t-t_1)/\tau_0)\right)^{1/\alpha}-\left(\ln(t^{(12)}/\tau_0)\right)^{1/\alpha}}{W^{(12)}}\right]\right\}$$

(25)

and $L_{12}(t) \Rightarrow 1-\exp\left[-\left(\frac{t}{\tau_1}\right)^{\beta_1}\right]-L_{1\bar{2}}(t).$  (26)

For the 60°-60°-60°-switching process, the switching probabilities defined in Section 2B after averaging read

$$L_{1\bar{1}}(t) \Rightarrow A \frac{\beta_1}{\tau_1} \int_0^t dt_1 \left(\frac{t_1}{\tau_1}\right)^{\beta_1-1} \exp\left[-\left(\frac{t_1}{\tau_1}\right)^{\beta_1}\right] \left\{\frac{1}{2} - \frac{1}{\pi}\arctan\left[\frac{\left(\ln((t-t_1)/\tau_0)\right)^{1/\alpha} - \left(\ln(t^{(11)}/\tau_0)\right)^{1/\alpha}}{W^{(11)}}\right]\right\},$$

(27)

$$L_{11\bar{1}}(t) \Rightarrow \frac{A^2}{\alpha\pi} \frac{\beta_1}{\tau_1} \int_0^t dt_1 \left(\frac{t_1}{\tau_1}\right)^{\beta_1-1} \exp\left[-\left(\frac{t_1}{\tau_1}\right)^{\beta_1}\right] \int_{\tau_0}^{t-t_1} d\tau \frac{\left(\ln(\tau/\tau_0)\right)^{1/\alpha-1} W^{(11)}/\tau}{\left[\left(\ln(\tau/\tau_0)\right)^{1/\alpha} - \left(\ln(t^{(11)}/\tau_0)\right)^{1/\alpha}\right]^2 + (W^{(11)})^2}$$

$$\times \left\{\frac{1}{2} - \frac{1}{\pi}\arctan\left[\frac{\left(\ln((t-t_1-\tau)/\tau_0)\right)^{1/\alpha} - \left(\ln(t^{(111)}/\tau_0)\right)^{1/\alpha}}{W^{(111)}}\right]\right\}$$

(28)

and  $L_{111}(t) \Rightarrow 1 - \exp\left[-\left(\frac{t}{\tau_1}\right)^{\beta_1}\right] - L_{1\bar{1}}(t) - L_{11\bar{1}}(t)$. (29)

The functions (20-29) will be implemented in the following in Eq. (16) to describe the polarization and strain response of a polycrystalline KNN material in the orthorhombic structure.

*D. Relation between polarization and strain variations*

A relation between the macroscopic strain, *s*, and the macroscopic polarization, *p*, can be derived from a general relation for electrostriction [53,54],

$$s_{ij} = Q_{ijmn} p_m p_n \tag{30}$$

where the electrostriction tensor $Q_{ijmn}$ is introduced and summation over the repeated indices implied. For ferroelectrics with a cubic parent phase, the piezoelectric coefficient is related to the spontaneous polarization *P* and can be derived from Eq. (30) if the spontaneous polarization singled out as

$$p_n \cong P_n + \varepsilon_0 \varepsilon_{nm} E_m \tag{31}$$

with the permittivity of vacuum $\varepsilon_0$ and the relative permittivity of the ferroelectric $\varepsilon_{nm}$. When substituting Eq. (31) into Eq. (30) the strain results as

$$s_{ij} \cong Q_{ijmn}P_m P_n + d_{ijk}E_k + \varepsilon_0^2 Q_{ijmn}\varepsilon_{ml}\varepsilon_{nk}E_l E_k \tag{32}$$

providing an expression for the piezoelectric coefficient [53]

$$d_{ijk} = 2\varepsilon_0 \varepsilon_{km} Q_{ijml} P_l . \tag{33}$$

The first term in Eq. (32) represents the spontaneous strain $S$. The last term in Eq. (32), quadratic in electric field, is by two orders of the magnitude smaller than the linear, piezoelectric term at highest fields involved in the experiment and thus will be neglected in the following.

We start with a single crystal problem, depicted in Fig. 1(a), and focus on the polarization and strain components in the direction of the applied field, *i.e.*, $[011]$. A corresponding strain variation may be calculated as the component $\Delta s'_{33}(t)$ in the new coordinate frame $(x', y', z')$, rotated about the initial *x*-axis in such a way that the $z'$ axis points in the field direction. The transformation matrix required for this rotation is

$$\mathbf{T} = \begin{pmatrix} 1 & 0 & 0 \\ 0 & 1/\sqrt{2} & -1/\sqrt{2} \\ 0 & 1/\sqrt{2} & 1/\sqrt{2} \end{pmatrix}. \tag{34}$$

The strain variation in the new frame reads $\Delta \mathbf{s}' = \mathbf{T}\Delta \mathbf{s}\mathbf{T}^T$ and the strain component of interest $\Delta s'_{33} = T_{3i}\Delta s_{ik}T^T_{k3} = (\Delta s_{22} + 2\Delta s_{23} + \Delta s_{33})$. In the "parent" coordinate frame of Fig. 1(a), collinear with the cubic parent phase, the tensor of electrostriction has a particularly simple form allowing the expression of all nonzero matrix elements through only three distinct matrix elements in Voigt notations, $Q_{11}, Q_{12}$, and $Q_{44}$ [54]. Thus, the spontaneous strain components

$$\Delta S_{22} = Q_{12}\left(P_1^2 + P_3^2\right) + Q_{11}P_2^2, \ \Delta S_{33} = Q_{12}\left(P_1^2 + P_2^2\right) + Q_{11}P_3^2, \ \Delta S_{23} = \Delta S_{32} = 2Q_{44}P_2P_3. \tag{35}$$

Now the time dependent strain variation along the field direction can be represented as

$$\Delta s(t) = \Delta S'(t) + \Delta s'_P(t), \tag{36}$$

where the first term represents a contribution of the spontaneous polarization to the strain (the first term in Eq. (32)) and the second one a contribution of the piezoelectric effect (the second term in Eq. (32)).

Considering the spontaneous strain contributions, we note that the strain variation during 90°-rotations, such as A-E and E-D switching steps, amounts to $\Delta S_{max} = -2Q_{44}P_s^2$ and $+2Q_{44}P_s^2$, respectively, denoting the maximum strain variation in a single switching event. During a 120°-rotations, such as A-C or B-D switching steps, the strain variation has an amplitude of $(Q_{11} - Q_{12} + 4Q_{44})P_s^2/4$. Since in the following we will study experimentally the average polarization and strain variations in polycrystalline materials, we may apply here the relation $2Q_{44} = Q_{11} - Q_{12}$ valid for ceramics [54]. This reduces the amplitude of the strain variation by 120°-rotations to $(3/4)\Delta S_{max}$. 60°-rotations contribute to the strain variation differently. Thus, in the triple 60°-sequence A-B-C-D, the first A-B step generates the strain variation of $(3/4)\Delta S_{max}$, the last C-D step the strain variation of $(-3/4)\Delta S_{max}$, while the switching step B-C does not change the strain. Generally, by all multistep switching events, the first step negatively contributes to the strain, while the consecutive steps contribute non-negatively, so that the complete variation at the end of the process equals zero. Taking all the contributions together, the spontaneous strain variation with time equals

$$\Delta S'(t) = \Delta S_{max} \left[ \eta_3 L_{3\bar{3}}(t) + \eta_2 \frac{3}{4} L_{2\bar{1}}(t) + \eta_{12} \frac{3}{4} L_{1\bar{2}}(t) + \eta_{11} \frac{3}{4} \left( L_{1\bar{1}}(t) + L_{11\bar{1}}(t) \right) \right]. \tag{37}$$

When calculating the piezoelectric strain contribution, we account at once for the fact that in poled ceramics, characterized by the Curie symmetry group $\infty m$, the dielectric permittivity tensor has the form $\hat{\varepsilon}' = \begin{pmatrix} \varepsilon_a & 0 & 0 \\ 0 & \varepsilon_a & 0 \\ 0 & 0 & \varepsilon_c \end{pmatrix}$ in the frame with the $z'$-axis oriented along the poling direction. Then the piezoelectric strain variation is transformed to

$$\Delta s'_P(t) = 2\varepsilon_0 \varepsilon_c Q_{11} E \left( \Delta P(t) - P_s \right), \tag{38}$$

when compared to the initial state without applied electric field ($E = 0$) and using the polarization variation along the field direction, Eq. (16) (for derivation see Appendix A).

The 180°- and the non-180° switching events contribute to the strain variation differently. The first ones only change the strain through the second term in Eq. (36) linear in polarization. In contrast, the non-180° switching processes rotate polarization by either 90°, 60° or 120°, and thus, change the strain by both terms in Eq. (36). We note also that the complete consecutive rotations along all paths cause no variation of the strain in the final state through the term quadratic in spontaneous polarization, Eq. (37), but they do through the term linear in spontaneous polarization, Eq. (38).

The above-developed MSM-model for single crystalline materials can also be applied to polycrystalline materials, although with some reservations. Crystalline grains of arbitrary shapes are randomly distributed and oriented in ceramics. For that reason, the maximum possible polarization of orthorhombic ceramics is limited by $0.912 P_s$ [55]. Another essential difference to a uniform medium is a nonhomogeneous distribution of the applied electric field, which is one of the reasons for the mentioned statistical distribution of local switching times accounted by the NLS [50,51] and the IFM [35,36] models. A combination of the MSM and the IFM models provided a better description of the time-dependent electromechanical response of tetragonal PZT ceramics [38]. Therefore, in the present work, we apply the hybrid MSM-IFM model, derived in Section 2C, together with the time-dependent expressions for strain (36-38) to the analysis of the polarization and strain response of the KNN-based ceramic during the polarization reversal. We note that both formulas for polarization, (16), and strain, (36), are assumed to be averaged over the whole polycrystalline system and neglect electric and elastic interactions [56] between different switching regions.

3. **Experimental details and sample characterization**

Bulk ceramics with chemical composition $(Na_{0.49}K_{0.49}Li_{0.02})(Nb_{0.8}Ta_{0.2})O_3$ and 2wt% $MnO_2$ addition were processed using a solid-state reaction route described elsewhere [57]. Green pellets were sintered at 1080 °C for 2 h and a density of 97.1% of the theoretical density was obtained. Sintered ceramic samples were ground to a thickness of 0.5 mm and sputtered with platinum electrodes, followed by 400 °C stress-relief annealing.

For structural analysis, sintered samples were crushed into powder and annealed at 500 °C for 2 h. X-ray diffractograms (XRD) were collected on the crushed powder using a diffractometer with Bragg-Brentano geometry (Bruker AXS D8 Advance, Germany) with Cu Kα radiation. The XRD of a NIST standard $LaB_6$ powder sample was measured to refine the instrumental profile. The Rietveld refinements were performed using the General Structure Analysis System (GSAS) software [58]. The starting parameters for the refinement were taken from the X-ray and neutron powder diffraction data refinements of Orayech *et al*. [59]. The refinement of structural parameters was restricted to the positions of Nb and O ions and the isotropic temperature factor, $U_{iso}$, was assumed uniform for the A-site, B-site, and oxygen ions, respectively.

X-ray powder diffraction profile of the ceramic sample is shown in Fig. 2, exhibiting typical perovskite structure without detection of secondary phase. An enlarged profile of the pseudo-cubic (220) and (002) reflections is shown in the inset. The intensity ratio of the (220) to (002) reflections is larger than unity, characteristic of an orthorhombic structure [60]. Rietveld refinement demonstrates that the pattern is well fitted with the orthorhombic *Amm*2 space group (see Supplementary materials). This is also confirmed by the temperature-dependent dielectric permittivity (Fig. 3), where the tetragonal to orthorhombic phase transition temperature is around 100 °C and thus, the phase structure is orthorhombic at room temperature.

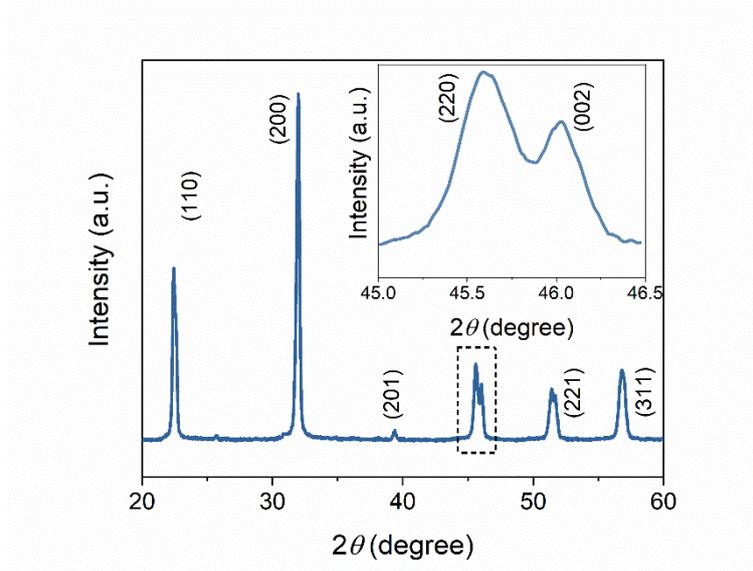

Figure 2. Powder XRD pattern of orthorhombic KNN sample. Reflections are indexed according to an orthorhombic setting. Enlarged profile of (220) and (002) reflections in the 2theta range of 44.5-47° is shown in the inset.

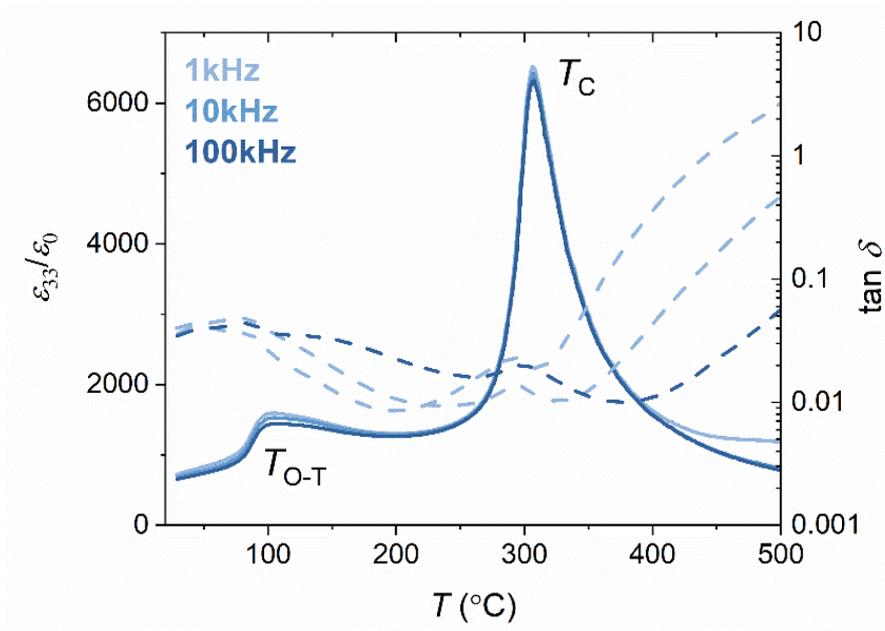

Fig. 3. Temperature-dependent dielectric constant and loss tangent of the investigated KNN ceramic sample, measured at 1 kHz, 10 kHz, and 100 kHz. Two dielectric anomalies are observed around 100 °C and 300 °C, corresponding to the orthorhombic-tetragonal phase transition temperature $T_{\text{O-T}}$ and the Curie temperature $T_{\text{C}}$, respectively. The ceramic material is in orthorhombic structure at room temperature.

Standard electromechanical characterization of the sample combining large-signal polarization and strain measurements is presented in Fig. 4. Large signal polarization hysteresis loops were measured using a modified Sawyer-Tower circuit and an electric field of 4 kV/mm with a frequency of 0.1 Hz was applied. The strain hysteresis loops were simultaneously obtained by measuring the displacement of the sample using an optical displacement sensor (D63, Philtec Inc., USA). The coercive field is 0.54 kV/mm. The maximum negative strain is -0.045% while the maximum positive strain is 0.086%.

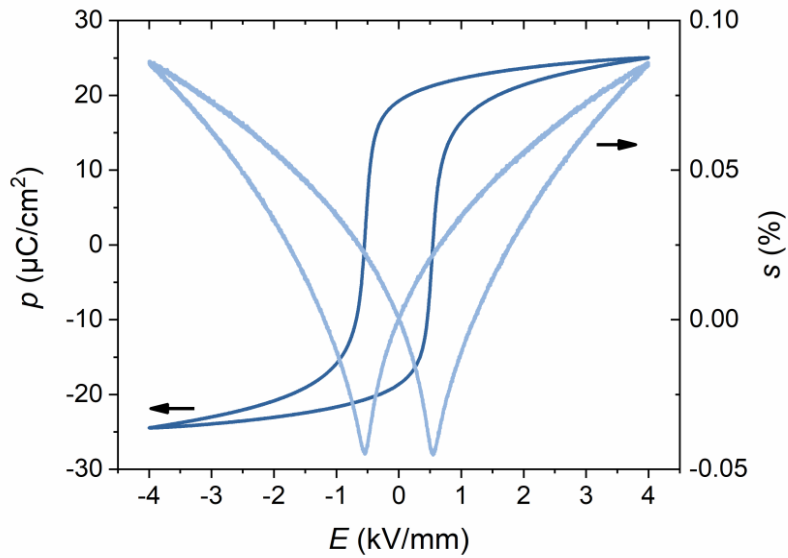

Fig. 4. Large signal bipolar polarization $p$ and strain $s$ loops, measured under 4 kV/mm and 0.1 Hz.

Time-dependent polarization reversal was measured using the pulse switching setup described in Ref. [31]. Each sample was first annealed and cycled with a bipolar electric field of 4 kV/mm at 0.5 Hz for 20 times to remove any potentially unstable electric contributions. Before each measurement, the samples were poled under 4 kV/mm for 10 s, followed by a relaxation time of 30 s to ensure the same starting conditions before each pulse application. A broad range of electric field pulses, $E_a$ (from $E_a = 0.15$ kV/mm to 1.30 kV/mm), which are approximately in the range of $0.3E_C < E_a < 2.6E_C$, were applied to the sample. In the meantime, the time dependence of the switched polarization and macroscopic strain response were simultaneously recorded. Bipolar polarization and strain hysteresis loops were measured before and after the pulse experiment to ensure it did not irreversibly affect the electrical properties. The full set of dynamical measurements may be found in Supplementary Materials. A set of representative switching curves is analyzed in the next section.

## 4. Model analysis of polarization and strain development over the time window

For the purpose of modelling, polarization and strain were measured over the time window for exemplary switching field values $E_a$ in the range 0.650-1.300 kV/mm where polarization reversal processes were completed. For the field values outside this range, only initial or final stages of the polarization and strain reversal were observed, which do not allow reliable evaluation of fitting parameters. Fig. 5 displays the time-dependent data (in red) together with theoretical fitting curves according to the MSM-IFM model, represented with black solid lines. Note that the leakage current and the reversible dielectric displacement were subtracted in the presented polarization $\Delta p$ data in Fig. 5.

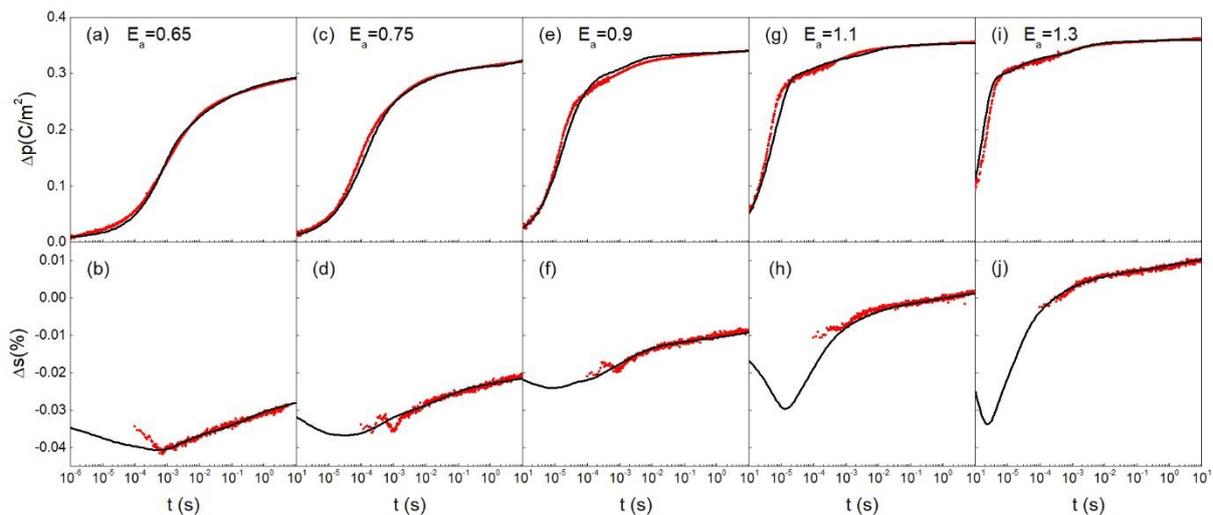

Fig. 5. Variation of the polarization (a,c,e,g,i) and strain (b,d,f,h,j) with time during field-induced polarization reversal at different applied field values in kV/mm, as indicated in the plots. Experimental curves and fitting curves are represented with the red symbols and black solid lines, respectively.

The variation of the strain starts in all measurements from zero (poled reference state), which is not explicitly seen in the plot since the data below $10^{-4}$ s are not experimentally accessible. We note a good quantitative agreement of the polarization and strain magnitudes achieved at t=10 s with the corresponding values at the respective fields in polarization and strain loops in Fig. 5. Because of accessible time window for polarization studies between $10^{-6}$ s and 10 s, the applied field range suitable for dynamic measurements appears to be far away from the polarization saturation achieved at about 4 kV/mm. The fact that not the whole system undergoes polarization reversal is reflected in the modelling by the adjustable field dependent values $P_s(E)$ and $\Delta S_{max}(E)$ as well as an adjustable vertical shift $\Delta S_0(E)$ added to strain, Eq. (36), to meet a proper strain value in the long time limit.

The KNN material studied here exhibits some dynamic polarization features different from tetragonal [8,34,35-37,50,51] and rhombohedral [8,32,38] PZT compositions and a tetragonal Cu-stabilized $0.94Bi_{0.5}Na_{0.5}TiO_3$–$0.06BaTiO_3$ (94BNT-6BT) material [36,61], which typically demonstrate a more or less smeared step-like behavior on the logarithmic time scale. Its behavior is also different from very dispersive response of $(1-x)Ba(Zr_{0.2}Ti_{0.8})O_3$-$x(Ba_{0.7}Ca_{0.3})TiO_3$ (BZT-BCT) compounds of different symmetries [9,10], which typically demonstrate almost straight-line dependence over many decades on the logarithmic time scale. In contrast, the KNN compound studied here reveals, at high fields above 0.9 kV/mm, two distinct initial temporal stages with different slopes (Fig. 4(e,g,i)) followed by a slow saturation at long times. This feature is reminiscent of the switching behavior of PZT composition at the morphotropic phase boundary [32], which might occur due to the phase coexistence. The KNN material studied here is, however, a single-phase one as is evidenced by X-ray data analysis (Fig. 2 and Fig. S1 in Supplementary materials) supported by dielectric data (Fig. 3). Another remarkable feature of the material is a rather dispersive strain reponse with a very small strain which is by one order of the magnitude smaller than that of tetragonal PZT [34] and shows comparatively minuscule changes through the time range, while polarization is about one half of that in this PZT.

To extract from the experiment a large number of parameters, involved in the theory, an educated guess based on the elementary Landau analysis is helpful (see Appendix B). To this end, free energy profiles were calculated for a chosen $K_{0.5}Na_{0.5}NbO_3$ solid solution, for which a set of verified Landau coefficients is known [62]. From the values of energy barriers and coercive fields for particular switching processes in the exemplary orthorhombic $K_{0.5}Na_{0.5}NbO_3$, the lowest activation fields $E_A$ may be expected for the 60°- and 90°-processes, while higher activation fields for the 120°- and the highest one for the 180°-process. Accordinly, it may be expected that the characteristic switching times of 60°-switching events, $\tau_1, t^{(11)}, t^{(111)}$, and of 90°-switching events, $\tau_3, t^{(33)}$, related to the activation fields $E_A^{(1)}, E_A^{(11)}, E_A^{(111)}$ and $E_A^{(3)}, E_A^{(33)}$, respectively, are much shorter than the characteristic times of 120°-switching events, $\tau_2, t^{(12)}$, related to the activation fields $E_A^{(2)}, E_A^{(12)}$. The latter characteristic times, in turn, are expected to be much shorter than that of the 180°-switching event, $t^{(4)}$, related to the activation field $E_A^{(4)}$.

Thus, it may be expected that fast switching events providing negative strain, occur at short times $\tau_1, t^{(11)}, t^{(111)}$ and $\tau_3, t^{(33)}$, which are mostly beyond the accessive time window for strain, while the consequent gradual increase of strain occurs at larger times $\tau_2, t^{(12)}$. From Eqs. (17c) and (23), the magnitude of the first switching step in the path 120°-60° appears to be proportional to $\tau_{111}/\tau_2 \ll 1$, and thus its contribution to the spontaneous strain is negligible. This makes the second swithing step in the process 60°-120° with the characteristic time $t^{(12)}$ presumably responsible for the late stage of the strain response observed at all exemplary applied fields.

The paradoxically low strain magnitude resulting from fast initial switching events, clearly visible in the polarization response, suggests that these processes do not contribute much to the spontaneous strain, as it is the case for the 180°-switching. However, the latter processes are characterized by high activation fields and expected in the best case at later switching stages. It is known, on the other hand, that fast non-180° processes may occur in a coherent way, so that their average macroscopic strain vanishes, and thus mimic the 180°-switching [63]. Such quasi-180° processes were indirectly identified in the electromechanical response of tetragonal PZT, with characteristic times typical of 90°-events [34,37], and of rhombohedral PZT, with characteristic times typical of 71°-events [38]. Very recently, such synchronized 90°-domain switching processes were directly observed by Zhu *et al.* during electric field loading of a barium titanate single crystal by using a fast camera combined with a polarized light microscope [64]. What coherent processes of this type are conceivable in orthorhombic ferroelectrics? Figure 6 displays possible coherent switching processes which do not cause macroscopic

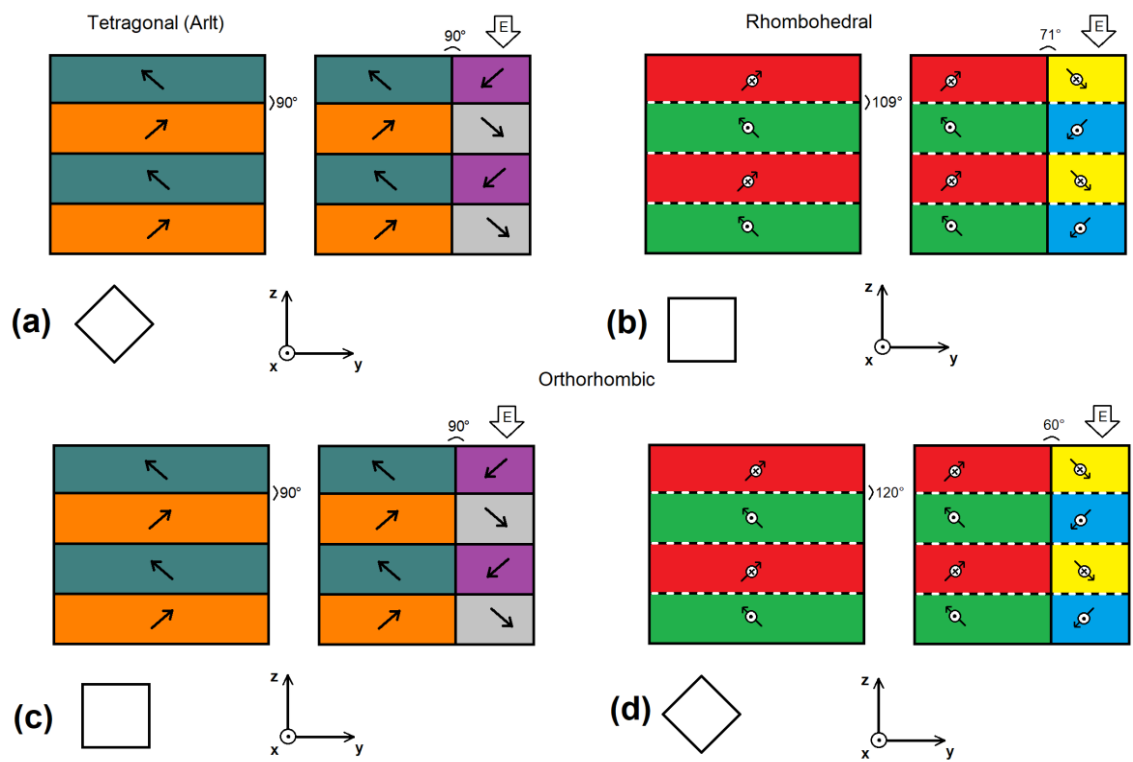

Fig. 6. (a) Array of equidistant horizontal 90°-domain walls in a tetragonal ferroelectric with a mean polarization in the [001]-direction is swept from right to left by a field-driven vertical 90°-domain wall that reverses the mean polarization. (b) A field-driven vertical 71°-domain wall (indicated by a solid line) sweeping through the equidistant 109°-domain wall system (indicated by dashed lines) and reversing the mean polarization in a rhombohedral ferroelectric. (c) Similar to the case (a) domain configuration and the sweeping domain wall in an orthorhombic ferroelectric. (d) A field-driven vertical 60°-domain wall (indicated by a solid line) sweeping through the equidistant 120°-domain wall system (indicated by dashed lines)

and reversing the mean polarization in an orthorhombic ferroelectric. Empty squares represent a pseudocubic unit cell orientation in each configuration.

spontaneous strain variation in ferroelectrics of different symmetries. As was suggested by Arlt [63], a 90°-domain wall driven by electric field and crossing a regular stack of 90°-domain walls in a tetragonal ferroelectric, as is shown in Fig. 6(a), does not cause a variation of macroscipic spontaneous strain because the strain changes in neighbor domains cancel each other. A fully analogous process is possible also in orthorhombic ferroelectrics as is shown in Fig. 6(c). The only difference consists in the domain wall orientation with respect to the pseudo-cubic unit cell which is rotated by 45°. A similar process in rhombohedral systems [38] is provided by a moving 71°-domain wall crossing a regular stack of 109°-domain walls as is shown in Fig. 6(b). The geometry of this configuration is a bit more complicated since the plane of a mechanically and electrically compatible 71°-domain wall is not perpendicular to the figure plane [65]. A process in orthorhombic systems analogous to the latter one is represented by a moving 60°-domain wall crossing a regular stack of 120°-domain walls as is shown in Fig. 6(d). Also here, the 120°-domain walls are rotated by 45° with respect to the pseudo-cubic unit cell orientation. Thus, in contrast to tetragonal and rhombohedral systems, orthorhombic ones allow two macroscopic strain-free non-180° switching scenarios.

To account for two coherent processes related to the sweeping 60°- and 90°-domain walls, respectively, we introduce additionally to the polarization response given by Eq. (16) two contributions, $\Delta P_{41}(t) = \eta_{41} 2 P_s L_{41}(t)$ and $\Delta P_{43}(t) = \eta_{43} 2 P_s L_{43}(t)$, where similar to Eq. (20)

$$L_{41}(t) = \frac{A}{\pi} \left\{ \arctan\left[ \frac{\left(\ln(t/\tau_0)\right)^{1/\alpha} - \left(\ln(t^{(41)}/\tau_0)\right)^{1/\alpha}}{W^{(41)}} \right] + \arctan\left[ \frac{\left(\ln(t^{(41)}/\tau_0)\right)^{1/\alpha}}{W^{(41)}} \right] \right\} \quad (39a)$$

and

$$L_{43}(t) = \frac{A}{\pi} \left\{ \arctan\left[ \frac{\left(\ln(t/\tau_0)\right)^{1/\alpha} - \left(\ln(t^{(43)}/\tau_0)\right)^{1/\alpha}}{W^{(43)}} \right] + \arctan\left[ \frac{\left(\ln(t^{(43)}/\tau_0)\right)^{1/\alpha}}{W^{(43)}} \right] \right\} \quad (39b)$$

with process fractions $\eta_{41}, \eta_{43}$, activation fields $E_A^{(41)} = E_A^{(1)}$, $E_A^{(43)} = E_A^{(3)}$, characteristic times

$$t^{(41)} = \tau_0 \exp\left[\left(\frac{E_A^{(1)}}{E_a(1+\kappa^2)}\right)^\alpha\right], \quad t^{(43)} = \tau_0 \exp\left[\left(\frac{E_A^{(3)}}{E_a(1+\kappa^2)}\right)^\alpha\right],$$

and width parameters

$$W^{(41)} = \frac{E_A^{(1)}}{E_a}\frac{\kappa}{1+\kappa^2}, \quad W^{(43)} = \frac{E_A^{(3)}}{E_a}\frac{\kappa}{1+\kappa^2}.$$

Note that the latter two coherent processes contribute to strain only through the total polarization in the piezoelectric part, Eq. (38).

By fitting polarization with $\Delta P(t) + \Delta P_{41}(t) + \Delta P_{43}(t)$ and strain with Eq. (36), using Eq. (16) and expressions (20-29) from the MSM-IFM model, the solid curves in Fig. 5 were obtained. A few parameters are known from independent measurements, like the relative permittivity $\varepsilon_c = 688$ and the electrostriction coefficient $Q_{11} = 0.0302$ m$^4$/C$^2$ derived from Fig. 4. The best fitting parameters common for all applied fields are $\tau_0 = 8\times10^{-12}$ s, as well as $\kappa = 0.15 \pm 0.01$ and $\alpha = 0.72 \pm 0.01$ which are close to the parameters of tetragonal PZT [37] and BNT-BT [61], respectively.

Table I presents activation fields, which determine characteristic times for all involved switching processes according to the assumed Merz law. As expected, the lowest values have the fields for the 90°- ($E_A^{(3)}, E_A^{(33)}$) and 60°-switching events ($E_A^{(1)}, E_A^{(11)}, E_A^{(111)} = E_A^{(21)}$), the medium values those for 120°-switchings ($E_A^{(2)}, E_A^{(12)}$) and the highest value that of 180°-switching, $E_A^{(4)}$. All activation fields remain stable over the studied applied field range supporting the validity of the Merz law for all processes.

Table I – Activation electric fields (kV/mm)

| $E_A^{(1)}$ | $E_A^{(11)}$ | $E_A^{(111)}$ | $E_A^{(12)}$ | $E_A^{(2)}$ | $E_A^{(21)}$ | $E_A^{(3)}$ | $E_A^{(33)}$ | $E_A^{(4)}$ |
|---|---|---|---|---|---|---|---|---|
| 42 | 29.5 | 23 | 55 | 56 | 23 | 36.5 | 27 | 75 |

Table II presents mechanical and electrical field-dependent fitting parameters as well as the Avrami indices. The maximum polarization $P_s$ and the magnitude of spontaneous strain $\Delta S_{max}$ show a monotonic rise with the increasing field while the strain shift $\Delta S_0$ a monotonic decrease of the magnitude compatible with the experimental data in Fig. 4. The Avrami indices $\beta_2$ and $\beta_3$ (where available) are not much different from unity. In contrast, the Avrami index $\beta_1$ is remarkably smaller than unity for the applied fields below $E$=0.9 kV/mm. Such a feature was previously observed for the first switching steps in tetragonal [34,37] and rhombohedral [38] ferroelectrics and is believed to be related to the support of the initial steps by residual stresses in the highly poled initial state [24]. The possibility of the index $\beta_1$ <1 is provided by a switching scenario II (see section 2B) where there are only latent nuclei but no new nucleations as suggested by Ishibashi *et al.* [21]. This implies also fractal geometries of nucleating domains with dimensionalities less than unity. It is noteworthy that $\beta_1$ exceeds unity for the fields higher than 0.9 kV/mm, for which the polarization curves become substantially different in shape (Fig. 5).

Table II – Avrami indices, mechanical parameters and spontaneous polarization

| $E_a$, kV/mm | $\beta_1$ | $\beta_2$ | $\beta_3$ | $\Delta S_{max}$, % | $\Delta S_0$, % | $P_s$, C/m² |
|---|---|---|---|---|---|---|
| 0.65 | 0.25 | - | 0.9 | -0.165 | -0.026 | 0.153 |
| 0.75 | 0.3 | - | 0.75 | -0.185 | -0.022 | 0.166 |
| 0.9 | 1 | 1 | 0.9 | -0.195 | -0.0123 | 0.1725 |
| 1.1 | 1.2 | - | 1 | -0.2 | -0.0026 | 0.179 |
| 1.3 | 1.5 | - | - | -0.205 | 0.0032 | 0.18 |

The fractions of different processes $\eta_i$ exhibit more complicated trends represented in Table III.

Table III – Process fractions

| $E$, kV/mm | $\eta_{11}$ | $\eta_{12}$ | $\eta_2$ | $\eta_3$ | $\eta_{41}$ | $\eta_{43}$ | $\eta_4$ |
|---|---|---|---|---|---|---|---|
| 0.65 | 0.17 | 0.08 | 0 | 0.4 | 0.35 | 0 | 0 |
| 0.75 | 0.17 | 0.05 | 0 | 0.4 | 0.28 | 0.1 | 0 |
| 0.9 | 0.17 | 0.05 | 0.05 | 0.4 | 0 | 0.33 | 0 |
| 1.1 | 0.51 | 0.15 | 0 | 0.135 | 0 | 0.145 | 0.06 |
| 1.3 | 0.73 | 0.2 | 0 | 0 | 0 | 0 | 0.07 |

These quantities are determined not only by activation barriers for particular processes, but also by the microstructure of the material like grain sizes and shapes, their orientation, properties of the grain boundaries, etc., and therefore may be considered as independent fitting parameters. The field dependence of these parameters appears to be distinct at low and high fields. The fractions of the switching paths 60°-60°-60° ($\eta_{11}$), 60°-120° ($\eta_{12}$) and 90°-90° ($\eta_3$) remain stable below $E=0.9$ kV/mm. At higher fields, $\eta_{11}$ and $\eta_{12}$ significantly increase, while $\eta_3$ significantly decrease and finally vanish. The coherent 60°-process with the fraction $\eta_{41}$ essentially contributes to switching below $E=0.9$ kV/mm but disappears at $E=0.9$ kV/mm and above. In contrast, the coherent 90°-process behaves non-monotonically; it is absent at both the lowest and the highest applied fields and its fraction $\eta_{43}$ reaches a maximum right at 0.9 kV/mm. We note that the two initial fast 90°- and 60°-switching processes, as well as their coherent counterparts, have characteristic switching times different by one order of the magnitude and therefore can be clearly resolved. Finally, the true 180°-switching process becomes visible only at the two highest applied fields and is quite essential for the adequate description of the long-time behavior in spite of relatively low fraction of $\eta_4=0.06$-$0.07$. The difference in process fractions at low and high fields reflects the respectively distinct shapes of the time-dependent polarization in these field regions. At the highest fields, the polarization reversal is strongly dominated by two processes, the faster 60°-60°-60°-switching and the slower 60°-120°-switching. We note that the true 180°-switching processes characterized by high activation fields and long switching times were not identified before in tetragonal [34,37] and rhombohedral [38] systems. Also noteworthy is a non-monotonic field dependence of the theoretical minimum strain value in Figs. 5(b,d,f,h,j) which is unfortunately beyond the region of reliably measured strain. This seems to be related with the change in the dominating switching scenario from the 90°-90° to the 60°-60°-60° path.

As a result, to describe the polarization switching of an orthorhombic ferroelectric/ferroelastic material, seven different possible switching paths should be taken into account (Fig. 7) which distinctly show themselves over the range of applied fields. The non-monotonous manner in which some of the fast processes appear and vanish with an increase of the applied field may hint at a competition between different mechanisms, which may appear simultaneously as several expanding domains, even within one grain, only to be devoured by the fastest growing domain. This might be the reason for only a brief appearance of the typically slow 120°-60° switching path at the applied field of 0.9 kV/mm around which the breakdown in the balance between 60°-60°-60° and 90°-90° switching processes also occurred. Residual stresses can also boost one of the fast processes giving it an edge over the other or even preclude one from happening if the strains it invokes cannot be realized due to, e.g., clamping with other grains.

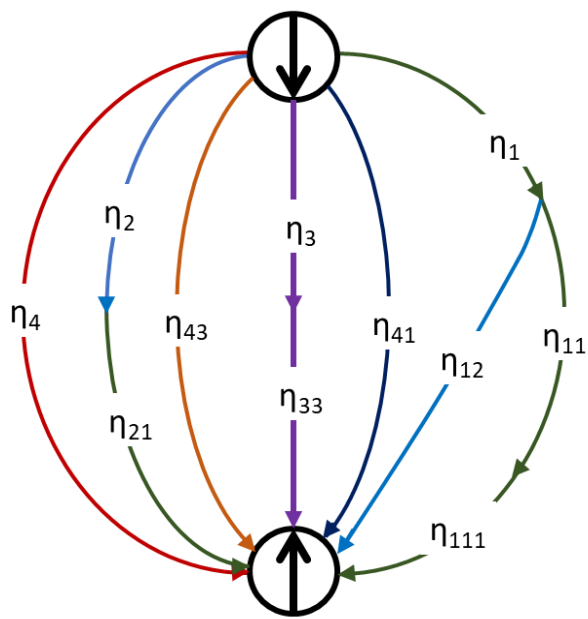

Fig. 7. A complete representation of all detected switching paths between the polar state marked with a down arrow to the polar state marked with an up arrow. In addition to the switching paths shown in the Fig. 1(b), two coherent switching processes as described via Fig. 7 were taken into account. The $\eta_{41}$ process is the 60° domain wall sweeping through the stack of 120° domains and $\eta_{43}$ depicts a process at which a 90° domain wall is sweeping through the 90° domains stack.

The 90°-processes could survive at higher fields only in a form of the coherent quasi-180° switching that does not produce strain. Ultimately, with the applied field increasing, the slow true-180° switching process makes a limited appearance. It is reasonable to hypothesise that at higher applied fields it will gain more prominence since when all potential barriers eventually flatten, this process would be the one not creating any strains and not requiring laminar domain structures to occur.

## 5. Conclusions

To understand the microscopic processes underlying the field driven polarization reversal in ferroelectrics/ferroelastics, it is critically important to investigate simultaneously electrical and mechanical responses to the applied electric field. Since particular switching processes may contribute exclusively or mostly to the polarization response, but do not contribute to the spontaneous strain, it appears possible to rectify contributions of particular switching channels, as it was done in the current study of the single-phase orthorhombic KNN ferroelectric ceramic.

The stochastic MSM-IFM model, extended in this work to the case of orthorhombic ferroelectrics/ferroelastics, allowed a thorough analysis of the original simultaneous measurements of polarization and strain response of orthorhombic KNN-piezoceramic over a wide time window. It enabled us to deconvolute the contributions of particular switching paths and quantify their fractions, activation fields, and Avrami indices, related to the dimensionality of the respective growing domains. Particularly, the analysis revealed a substantial contribution of the fast processes, which do not cause macroscopic spontaneous strain and exhibit characteristics of non-180° switching events. Such coherent non-180° switching processes were indirectly identified in previous studies [34,37,38] and recently directly observed optically [64]. This is in agreement with a general conclusion from molecular dynamics that 180°-switching in tetragonal ceramics proceeds rather via sequential 90°-domain wall motion [25]. Coherent processes may also be related to the recently observed avalanche switching behavior in such diverse ferroelectric/ferroelastic systems as $Pb(Mg_{1/3}Nb_{2/3})O_3$–$PbTiO_3$ and $BaTiO_3$ with quite distinct domain shapes and structures [18].

The advanced approach should, however, be further elaborated. The main limitation of the actual MSM-IFM model lies in the assumed Lorenz statistical distribution of the local electric fields, which substantially reduces the flexibility of the model. In fact, statistical field distributions observed in different materials may be quite asymmetrical [8-10,36,61,66]. In some materials, like HfO$_2$-based ferroelectrics, Gaussian proved to be more suitable for description of the statistical field distribution [66-68]. Using a more adequate statistical field distribution may further improve description of the dynamic electromechanical response of ferroelectrics/ferroelastics.

**Acknowledgements**

This work was supported by the Deutsche Forschungsgemeinschaft (DFG) Grants Nos. 405631895 (GE-1171/8-1 and KO 5100/1-1) and GR 479/2. K.W. acknowledges the support of the National Nature Science Foundation of China (Grant Nos. 51822206 and 52032005)

**Appendix A. Relation of polarization and piezoelectric strain for orthorhombic symmetry**

In a single crystal, the *piezoelectric* contribution to the strain along the poling direction $[011]$, calculated in the new coordinate frame $(x', y', z')$, given by $\mathbf{s'} = \mathbf{T s T}^T$ with Eq. (34), equals

$$s_{33}'^P = \frac{\varepsilon_0 E}{\sqrt{2}} \left\{ \begin{array}{l} 2Q_{12}(\varepsilon_{21} + \varepsilon_{31})P_1 + \left[(Q_{11} + Q_{12})(\varepsilon_{22} + \varepsilon_{32}) + 2Q_{44}(\varepsilon_{23} + \varepsilon_{33})\right]P_2 \\ \left[(Q_{11} + Q_{12})(\varepsilon_{23} + \varepsilon_{33}) + 2Q_{44}(\varepsilon_{22} + \varepsilon_{32})\right]P_3 \end{array} \right\}. \quad \text{(A1)}$$

Considering the diagonal form of the permittivity tensor in the new frame with the $z'$-axis oriented along the poling direction, which implies the Curie symmetry group $\infty m$ as for poled ceramics [54], $\hat{\varepsilon}' = \begin{pmatrix} \varepsilon_a & 0 & 0 \\ 0 & \varepsilon_a & 0 \\ 0 & 0 & \varepsilon_c \end{pmatrix}$ the permittivity tensor in the initial frame becomes

$\hat{\varepsilon} = \begin{pmatrix} \varepsilon_a & 0 & 0 \\ 0 & \frac{\varepsilon_c + \varepsilon_a}{2} & \frac{\varepsilon_c - \varepsilon_a}{2} \\ 0 & \frac{\varepsilon_c - \varepsilon_a}{2} & \frac{\varepsilon_c + \varepsilon_a}{2} \end{pmatrix}$. This reduces the formula (A1) to

$$s'^{P}_{33} = \frac{\varepsilon_0 \varepsilon_c E}{\sqrt{2}}(Q_{11} + Q_{12} + 2Q_{44})(P_2 + P_3). \tag{A2}$$

From $\mathbf{P}' = \mathbf{TP}$ it follows that $P'_3 = (P_2 + P_3)/\sqrt{2}$. Using further the relation $2Q_{44} = Q_{11} - Q_{12}$ valid for ceramics one finds

$$s'^{P}_{33} = 2\varepsilon_0 \varepsilon_c Q_{11} E P'_3. \tag{A3}$$

When comparing with the initial state without an applied electric field but after poling in $[0\bar{1}\bar{1}]$ direction, the polarization variation $\Delta P(t) = P'_3(t) - (-P_s)$ which results in the formula for strain variation, Eq. (38).

**Appendix B. Elementary Landau analysis of switching barriers**

An elaborated Ginzburg-Landau-Devonshire (LGD) model for the particular KNN material under consideration is not available. However, a qualitative field behaviour of the switching barriers can be studied by using the data for similar orthorhombic KNN compounds [62]. Thus, an 8[th] order Landau free energy density as function of polarization components reads

$$G_L = \alpha_1\left(P_1^2 + P_2^2 + P_3^2\right) + \alpha_{11}\left(P_1^4 + P_2^4 + P_3^4\right) + \alpha_{12}\left(P_1^2 P_2^2 + P_1^2 P_3^2 + P_2^2 P_3^2\right) + \alpha_{123} P_1^2 P_2^2 P_3^2 + \alpha_{111}\left(P_1^6 + P_2^6 + P_3^6\right)$$
$$+ \alpha_{112}\left[P_1^2\left(P_2^4 + P_3^4\right) + P_2^2\left(P_1^4 + P_3^4\right) + P_3^2\left(P_1^4 + P_2^4\right)\right] + \alpha_{1111}\left(P_1^8 + P_2^8 + P_3^8\right) + \alpha_{1122}\left(P_1^4 P_2^4 + P_1^4 P_3^4 + P_2^4 P_3^4\right)$$
$$+ \alpha_{1112}\left[P_1^6\left(P_2^2 + P_3^2\right) + P_2^6\left(P_1^2 + P_3^2\right) + P_3^6\left(P_1^2 + P_2^2\right)\right] + \alpha_{1123}\left(P_1^4 P_2^2 P_3^2 + P_1^2 P_2^4 P_3^2 + P_1^2 P_2^2 P_3^4\right) - E_1 P_1 - E_2 P_2 - E_3 P_3,$$

(B1)

where the Landau coefficients $\alpha$ for the orthorhombic compound $K_{0.5}Na_{0.5}NbO_3$ at temperature T=300 K can be found by Pohlmann et al. [62]. The highest energy barrier for switching between different stable polarization states in the absence of external electric field appears to be that for the direct 180°-polarization reversal (e.g. AD path in Fig. 1(a)), while the lowest barrier exists for any 60°-polarization switching (e.g. AB path in Fig. 1(a)). The relation between barriers for 180°-, 120°-, 90°- and 60°-processes is $\Delta G_{180} : \Delta G_{120} : \Delta G_{90} : \Delta G_{60} = 7.49 : 4.18 : 1.76 : 1$. When electric field $E$ is applied in $[011]$ direction, the energy profile (B1) is tilted and the mutual relations between barriers substantially change because the field has different projections at different switching paths. That is why the coercive fields for overcoming these barriers exhibit another sequence. Thus the barrier for the second 60°-process (e.g. BC path in Fig. 1(a)) disappears first at the coercive field of $E = 1.1 \times 10^7$ V/m. In the course of further field increasing the next barriers to vanish at the field $E = 2.14 \times 10^7$ V/m are those for both 90°-processes in the 90°-90° switching path (e.g. AE and ED paths in Fig. 1(a)). The next barrier disappearing at a bit larger field $E = 2.18 \times 10^7$ V/m is the one for the third 60°-process (e.g. CD path in Fig. 1(a)). The barrier for the first 60°-process (e.g. AB path in Fig. 1(a)) vanishes at the field $E = 2.25 \times 10^7$ V/m. A substantially larger field $E = 3.22 \times 10^7$ V/m is required to suppress the barriers for both 120°-switching steps (e.g. AC and BD paths in Fig. 1(a)) and a further enhanced field $E = 4.13 \times 10^7$ V/m to suppress the barrier for the 180°-switching (AD path in Fig. 1(a)). We note that the above relations between the energy barriers and coercive fields result from a very simple Landau analysis and can be remarkably modified by ferroelastic interactions, particularly, by residual stresses in highly poled states.